# On PMU Adequacy for Wideband Dynamics Tracking Beyond Attenuation Assumptions

Chetan Mishra, *Member, IEEE*, Luigi Vanfretti, *Senior Member, IEEE,* Jaime Delaree Jr., *Member, IEEE* and, Kevin D. Jones, *Senior Member, IEEE*

***Abstract*— Recent work highlights the inadequacy of standard synchrophasors in capturing high-frequency grid dynamics, including data center oscillations, primarily due to filtering and windowing attenuation during phasor estimation. This letter shows how synchrophasor ambient data spectrum itself contains information on attenuation, which provides a way to identify critically impacted measurements and potentially estimate and undo the effect. Results are shown for synthetic data as well as actual synchrophasor estimates from digital fault recorders (DFRs) within the Dominion Energy system.**

## I. Introduction

Synchrophasors are critical for tracking power grid dynamics, particularly with the rise of unforeseen wideband oscillations due to the growing integration of power electronic-based devices, which often lack accurate models to analyze such oscillations. However, conventional signal processing chains in synchrophasor estimation algorithms limit tracking capabilities by attenuating oscillation magnitudes as well as introducing phase shifts [1]. This may impact oscillation monitoring applications, which rely on relative magnitudes across devices to localize oscillations and quantify severity [2]. On the other hand, signals measured by a single device have the same phase distortion [1] so applications relying on relative angles (e.g., P, Q) remain robust to it. Current literature [1] [2] [3] examines this impact using phasor measurement unit (PMU) data from grid disturbances and oscillations. However, this work primarily attributes attenuation to loss of periodic-oscillation information, implying a lower signal-to-noise ratio (SNR) that limits synchrophasors' ability to analyze fast behavior. Left unaddressed, this incomplete picture risks leaving PMU owners with the impression that their devices are unsuited for tracking such dynamics, an impression this letter shows to be unwarranted. In this work, we derive the impact of attenuation on ambient noise and show that it preserves SNR, thereby directly supporting the case that PMUs remain adequate for wideband dynamics tracking, for example capturing a 22 Hz, low magnitude oscillation in [4]. This also provides a way to estimate (and potentially correct) attenuation directly from the synchrophasor ambient spectrum, at the very least helping system operators identify and prioritize affected devices.

## II. Main Result

Power system voltage/current analog signal $x(t)$ can be modeled in the following general form,

$$x(t) = \sum_{i\geq 1} a_i(t) e^{j2\pi f_i t} \quad (1)$$

where the $i^{th}$ term represents $f_i$ frequency component, with complex amplitude $a_i(t) = A_i(t)e^{j\phi_i(t)}$ , where $A_i(t) > 0$ and $\phi_i(t)$ are magnitude and phase respectively, with $a_i(t)$ having a dominant DC component, i.e. $|\hat{a}_i(0)| \gg |\hat{a}_i(f \neq 0)|$. Here, $\hat{x}(f)$ represents Fourier transform of arbitrary signal $x(t)$. The fundamental component (50 or 60 Hz) is denoted by $i = 1$ i.e. at $f_1$ Hz, which is what the synchrophasor estimate aims to extract. The next subsection examines how practical DFT-based estimation of this component departs from this ideal model, and what that means for tracking wideband dynamics.

### A. Sources of Error in DFT Based Synchrophasor

A typical synchrophasor algorithm [1] samples the analog voltage/current signal $x(t)$ followed by a moving window Discrete Fourier Transform (DFT) to extract its fundamental component, followed by filtering and downsampling to a specified reporting rate ( e.g., 30, 60 of 120 samples per sec.).

Mathematically, the DFT used in synchrophasors is the Short-Term Fourier Transform (STFT) evaluated at a single frequency of $f_1$ Hz. The STFT coefficient $V_g x(t,f)$ of $x$ at time $t$ and frequency $f$ using a real windowing function $g(t)$ is defined as [5],

$$V_g x(t,f) = \int x(\tau) g(\tau - t) e^{-j2\pi f \tau} d\tau. \quad (2)$$

Here, most standard windows satisfy $\int g(\tau)d\tau = 1$ and $\lim_{f\to 0}|\hat{g}(f)| \to 1$. To get the synchrophasor estimate $y(t)$ of $a_1(t)$, we substitute $x(t)$ into expression above, evaluate at $f_1$ Hz,

$$y(t) = V_g x(t, f_1) = \int a_1(\tau) g(\tau - t) d\tau + \sum_{i>1} \int a_i(\tau) e^{j2\pi(f_i - f_1)\tau} g(\tau - t) d\tau \quad (3)$$

$$\Rightarrow \hat{y}(f) = \underbrace{\hat{g}(-f)\hat{a}_1(f)}_{\substack{attenuation\ of \\ fund.\ dynamics}} + \underbrace{\hat{g}(-f)\sum_{i>1} \hat{a}_i(f + f_1 - f_i)}_{\substack{inter\ harmonic \\ interference}} .$$

The first term above shows that $f$ Hz component of $f_1$ suffers attenuation by $|\hat{g}(-f)| \leq 1$ and phase distortion by $\angle\, \hat{g}(-f)$. Note that, as mentioned before, $y(t)$ can be further

Chetan Mishra, Jaime Delaree Jr. and Kevin D. Jones are with Dominion Energy, Richmond, VA 23059 USA (e-mail: chetan31@vt.edu).
Luigi Vanfretti is with Rensselaer Polytechnic Institute, Troy, NY, USA (e-mail:vanfrl@rpi.edu )

filtered before downsampling for reporting, in which case we simply replace $g(t)$ with $(g * h)(t)$ and $\hat{g}(-f)$ by $\hat{g}(-f)\hat{h}(f)$ in the above expressions, where $*$ is convolution operator and $h$ is the filter kernel with frequency response $\hat{h}(f)$. Minimizing this error requires $\hat{g}(f) \approx 1$ in a broad range. Note that a short window (in time) is highly desirable for a high reporting rate and minimizes latency in real-time control applications, as done in P class PMUs [2], where 1-2 cycle window is used, which forces $\hat{g}(f)$ to have a bigger spread in frequency domain owing to *Heisenberg Uncertainty Principle* [5], thereby being desirable for minimizing the attenuation above.

However, doing so also introduces contributions from other frequencies $f_i \neq f_1$, (harmonics and inter-harmonics), which appear as sinusoidal terms at frequencies $\{f_i - f_1 \forall i > 1\}$. This follows from substituting $\hat{a}_i(f) = \hat{a}_i(0)\delta(f)\forall i > 1$ into the second (error) term of (3):

$$y(t) = fundamental + \underbrace{\textstyle\sum_{i>1} \hat{g}(f_1 - f_i)\hat{a}_i(0)e^{j2\pi(f_i - f_1)t}}_{\substack{inter\ harmonics \\ as\ observed\ oscillations}}. \quad (4)$$

Thus, the closer an inter-harmonic $f_i$ is to $f_1$, the larger $|\hat{g}(f_1 - f_i)|$ and the more prominent its impact on the final phasor estimate, potentially being mistaken for oscillations [6]. Controlling this error requires $|\hat{g}(f)|$ to drop off sharply, as done in M class PMUs, which use longer windows (4–6 cycles) for monitoring applications. This, however, can attenuate genuine higher-frequency components near $f_1$ Hz. This assumption relies on power system dynamics being relatively slow, i.e., $\hat{a}_1(f) \approx 0 \forall f > 2\ Hz$, allowing an aggressive roll off in $|\hat{g}(f)|$ (using a longer time-domain window $g(t)$) can be applied without attenuating genuine dynamics.

### *B. Attenuation Effect in Synchrophasor Ambient Spectrum*

In real world power systems, $A_1(t) = T_A(t) + \epsilon_A(t)$ and $\phi_1(t) = T_\phi(t) + \epsilon_\phi(t)$, where $T\ (t)$ represents the DC component plus a slow trend ($< 0.1\ Hz$) and $\epsilon\ \ (t)$ is ambient dynamics and measurement noise, modeled as a zero-mean stationary stochastic component [7]. For a short synchrophasor window, pulling the trend terms outside the integral simplifies (3) as,

$$y(t) \approx e^{jT_\phi(t)} \int \big(T_A(t) + \epsilon_A(\tau)\big)e^{j\epsilon_\phi(\tau)}g(\tau - t)d\tau. \quad (5)$$

For small enough ambient dynamics $\epsilon_\phi(t), \epsilon_A(t)$, taking linear approximation $e^{j\epsilon_\phi(\tau)} \approx 1 + j\epsilon_\phi(\tau)$ and neglecting second order terms in $\epsilon$ gives

$$y(t) \approx e^{jT_\phi(t)} \Big(T_A(t) + \int \epsilon_A(\tau)g(\tau - t)d\tau + jT_A(t) \int \epsilon_\phi(\tau)g(\tau - t)d\tau\Big). \quad (6)$$

Taking the magnitude $|y(t)|$ and ignoring the quadrature term (second order in $\epsilon_\phi(\tau)$) yields,

$$|y(t)| \approx T_A(t) + \int \epsilon_A(\tau)g(\tau - t)d\tau. \quad (7)$$

Finally, detrending yields the following relationship between power spectral density (PSD) of synchrophasor magnitude spectrum $S_y(f)$ and spectrum of $\epsilon_A$ denoted by $S_A(f)$,

$$S_y(f) = |\hat{g}(f)|^2 S_A(f) \quad (8)$$

### *C. Recovering (Partial) Information About Attenuation*

From (3), given full knowledge of $g(t)$ (consequently $\hat{g}(f)$), one can mitigate errors in both magnitude and phase at each frequency $f$ in $a_1(t)$. This information is not usually available; however, it can be estimated in a lab testing environment by injecting a known signal e.g. $x(t) = \underbrace{e^{j2\pi f_{osc}t}}_{a_1(t)} \times e^{j2\pi f_1 t} \Rightarrow \hat{a}_1(f) = \delta(f - f_{osc})$ into the PMU device, where the $f_{osc}$ component in the output i.e. $\hat{y}\ (f_{osc}) = \hat{g}(-f_{osc})\delta(f - f_{osc})$, which from (3) directly yields $\hat{g}(-f_{osc})$). However, this is not feasible in practice since, besides being labor intensive, it requires taking each device out of service. Moreover, firmware upgrades can change this response, requiring re-testing.

Equation (8) shows that the estimation window attenuates the true amplitude spectrum $S_A(f)$. This attenuation $|\hat{g}(f)|$ cannot be estimated directly because $S_A(f)$ is still unknown. For a hypothetical white noise process $\epsilon_A(t)$, $S_A(f) = \sigma^2$ (constant) [7] $\Rightarrow \hat{S}_y(f) \propto |\hat{g}(-f)|^2$ so the measured magnitude spectrum, together with a standard window condition on $g$ such as $\lim_{f \to 0} |\hat{g}(f)| \to 1$ can help estimate $|\hat{g}(-f)|$. However, $S_A(f)$ reflects the grid's response to random perturbations and is not a constant, which is also the basis for output-only modal analysis using ambient data. Real-world measurements are further corrupted by thermal and quantization noise from current and potential transformers (CTs/PTs), which results in white noise entering upstream at the waveform stage. As a result, $S_A(f) = S_A^{dyn}(f) + S_A^{noise}(f)$, where the noise power $S_A^{noise}(f) \approx \sigma^2$ forms a relatively flat noise floor. Critically, this "attenuation" does not affect the signal-to-noise ratio itself, thereby limiting its impact on synchrophasor-based detection applications. Based on our measurements, $S_A(f)$ is dominated by $S_A^{dyn}(f)$ in low frequency ranges (particularly < 2 Hz), while higher frequency ranges are more dominated by $S_A^{noise}(f)$, and therefore relatively flat. This allows us to take the following assumption,

**Assumption 1:** $\exists f_h, \sigma\ s.t.\ S_A(f) \approx \sigma^2\ a.e.\ in\ (f_h, \infty)$.

Consequently, $S_y(f) \approx \sigma^2 |\hat{g}(f)|^2 \forall f > f_h$, allowing us to recover a *scaled* version of the attenuation function at higher frequencies. Note that the qualifier "almost everywhere" (a.e.) allows $S_A(f)$ at a countable set of frequencies $f$ e.g., oscillations, which have discrete spectra [7] and can be handled as outliers in the downstream estimation process. However, the larger the chosen $f_h$ value, the more likely Assumption 1 holds at the expense of essential information on the attenuation level contained in the *main lobe* of the window function $\hat{g}(f)$. Next, we quantify attenuation via the slope $\beta$ of a log-linear fit of $S_y(f)$ vs $f$ (from Assumption 1 and (8)),

$$10\log_{10}\big(S_y(f)\big) \approx 10\log_{10}(\sigma^2) + 20\log_{10}(|\hat{g}(f)|) \approx C - \beta f\ \forall f \in \big(f_h, f_{nyq}\big) \quad (9)$$

where $f_{nyq}$ is the Nyquist frequency. This metric can be used to rank devices for lab testing and is also robust to the ambient noise power $\sigma^2$ itself, enabling comparison across locations. In practice, Assumption 1 may hold only approximately, since the true spectrum is rarely perfectly constant, this leaks into the $\beta$ estimate. So what counts as an abnormal value depends on system properties. Regardless, severely affected PMU devices will appear as outliers.

## III. Results

### A. Synthetic Data

Here, we generate 60 Hz waveform data (Figure 1 a)) with fixed phase $\phi_1(t) = 0$ , modulated by an average 1-unit amplitude $A_1(t)$ comprised of a slow trend (Figure 1 b)), two well damped modes at 1 and 2 Hz as well as a small 10 Hz oscillation (visible in magnitude $S_A(f)$ in Figure 2). Next, varying levels of 0-mean Gaussian noise are added to the waveform data, with the measured amplitude spectra shown in Figure 2. In this case, the measurement noise "submerges" the true spectrum to the extent that dynamics with power lower than the measurement noise become unidentifiable, resulting in a flat measured amplitude spectrum i.e. dominated by noise power $S_A^{noise}(f)$. In contrast, the true, noise-free spectrum consists purely of $S_A^{dyn}(f)$ and therefore continues to decline at higher frequencies owing to absence of fast dynamics. Thus, the lower the SNR, the more "flattened" the magnitude spectrum entering the synchrophasor estimation stage becomes.

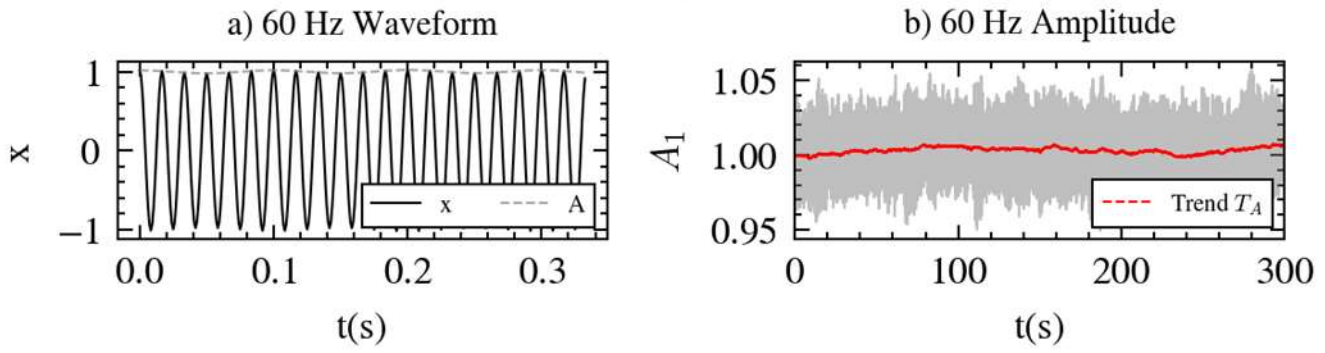

**Figure 1. Synthetic Data**

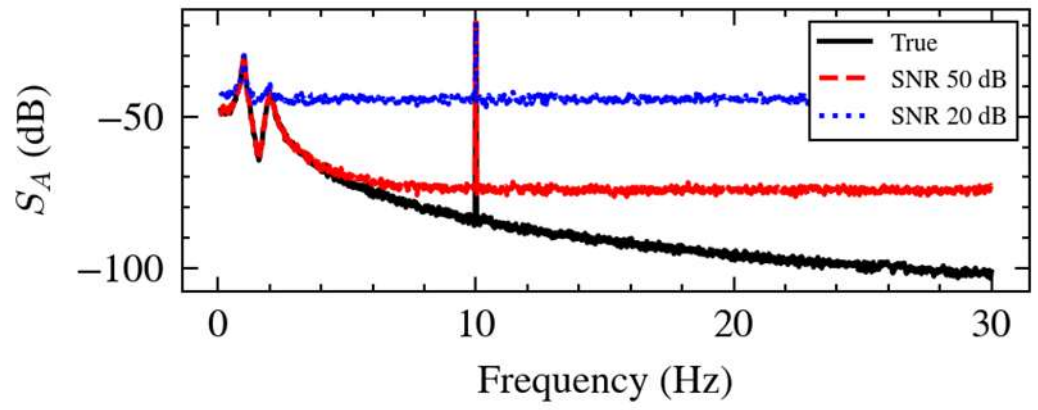

**Figure 2. Impact of Measurement Noise on Synchrophasor Spectrum**

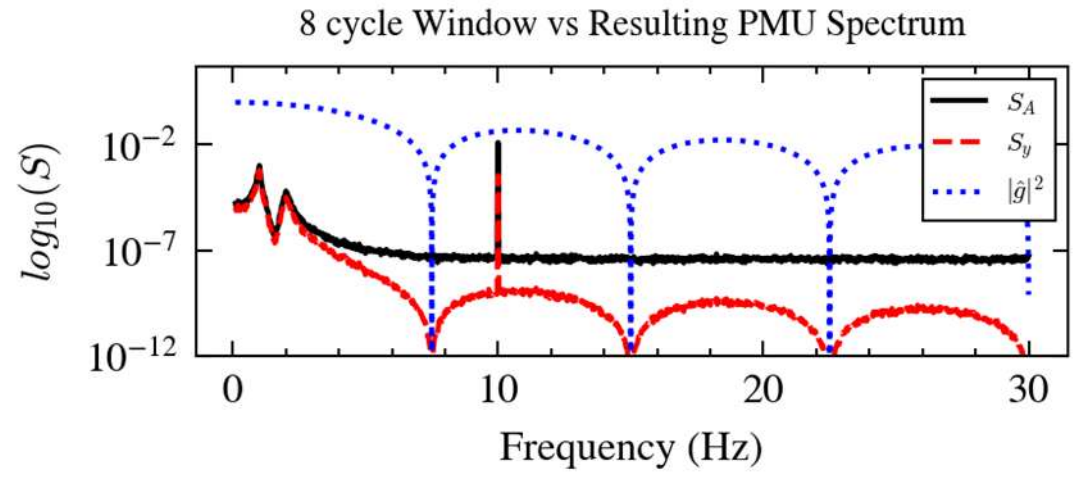

**Figure 3. Window Response Observable in Synchrophasor Spectrum**

Proceeding with the 50 dB SNR waveform data, which largely retains prominent dynamics, we can see that $f_h \approx 5$ Hz can serve as a good value for satisfying Assumption 1. Consequently, the resulting synchrophasor data magnitude log spectrum in Figure 3 (in red) for 8 cycle standard boxcar window $\forall f \geq 5\,Hz$ has similar shape as the gain response / attenuation function $|\hat{g}(f)|$ (in blue) of the underlying window due to (8). In this example, with prior knowledge of window type (boxcar), the entire window response $|\hat{g}(f)|$ can be estimated by using the positions of zeroes (integer multiples of $\frac{60}{n_{cycle}}$ Hz) in the observed (partial) response in $S_y(f > f_h)$.

### B. Dominion Energy Data

Figure 4 shows a PSD estimate (zoomed to 30–90 Hz) from a 3-minute window of point-on-wave voltage data recorded by a digital fault recorder (DFR) sampling at 960 Hz, at a 115 kV bus of a solar PV plant. One prominent feature is a pair of sidebands at 82 and 38 Hz from a 22 Hz mode analyzed in [4].

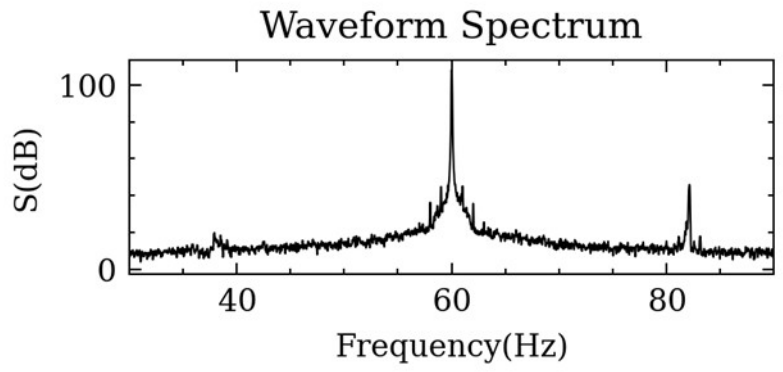

**Figure 4. Waveform Spectrum**

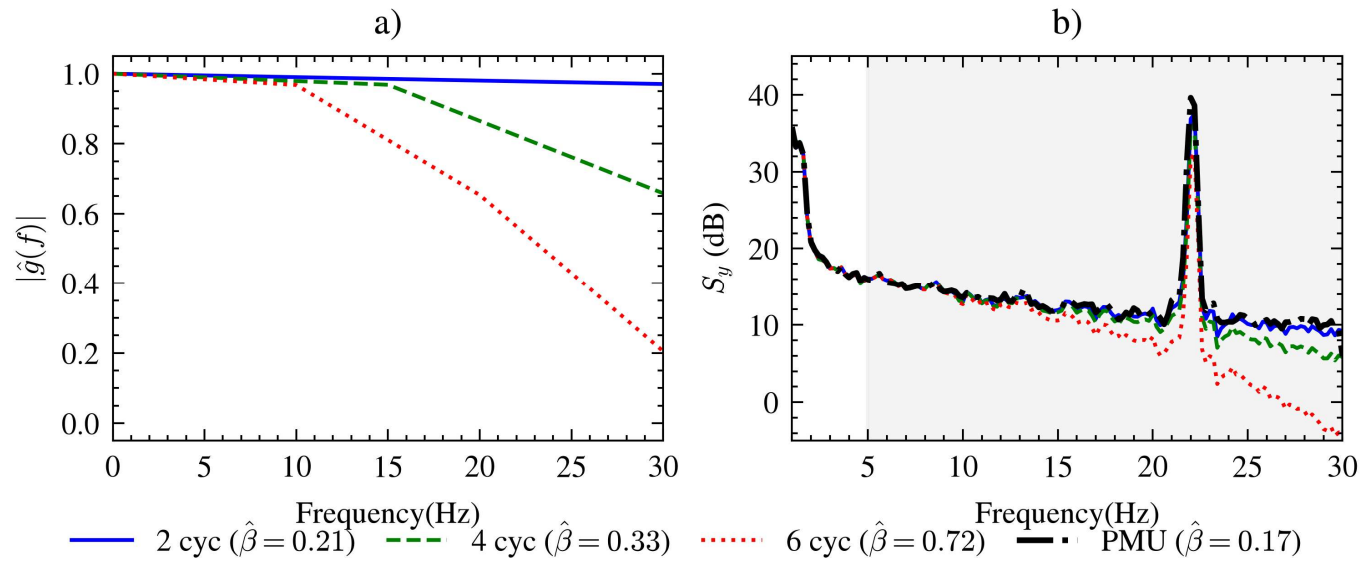

**Figure 5. a) Flattop window frequency response vs window length b) resulting PMU estimate spectra along 1with DFR estimate**

Next, we analyze the synchrophasor estimate from the same device, which is reported at 60 Hz. Since we do not know the ground truth synchrophasor data, in Figure 5 b), we benchmark it against DFT-based synchrophasor estimates using a Flattop window, which has the least attenuation among windows of the same length in the time domain. We use different window lengths to assess the quality of the reported synchrophasor estimate, which shows very low attenuation, even better than a 2-cycle estimate, which already has negligible attenuation up to 30 Hz as shown in Figure 5 a). Furthermore, with $f_h = 5$ Hz (avoiding the dominant dynamics below 2 Hz), the proposed metric $\hat{\beta}$ (attenuation rate in dB/Hz) increases with longer windows, potentially providing a way to quantify the level of attenuation without the knowledge of ground truth.